\documentclass[useAMS,usenatbib]{mn2e}
\usepackage{epsfig}
\usepackage{rotating}
\usepackage{times}
\usepackage{graphicx}
\usepackage{lscape}
\usepackage{aas_macros,graphicx,times,multirow}

\def \sw {{\em Swift}} 

\begin{document}
\title[Low-Luminosity AGNs with \sw]{Variability and spectral energy 
distributions of low-luminosity active galactic nuclei: a simultaneous X-ray/UV look with \sw}
\author[E.\ Pian et al.]{E.\ Pian$^{1,2,3}$\thanks{E-mail: elena.pian@sns.it},
P.\ Romano$^{4}$,  D.\ Maoz$^{5}$,  A.\ Cucchiara$^{6}$,  C.\ Pagani$^{6}$, and V.\ La Parola$^{4}$ \\
$^{1}$ INAF, Osservatorio Astronomico di Trieste, Via G. Tiepolo 11,  I-34143  Trieste, Italy\\
$^{2}$ Scuola Normale Superiore, Piazza dei Cavalieri 7, I-56126 Pisa, Italy\\
$^{3}$ European Southern Observatory, Karl-Schwarzschild-Strasse 2, D-85748 Garching bei M\"unchen, Germany \\
$^{4}$ INAF, Istituto di Astrofisica Spaziale e Fisica Cosmica, Via U.\ La Malfa 153, I-90146 Palermo, Italy\\
$^{5}$ School of Physics and Astronomy, Tel-Aviv University, Tel-Aviv 69978, Israel \\
$^{6}$ Department of Astronomy and Astrophysics, Pennsylvania State University, 
525 Davey Laboratory, University Park, PA 16802, USA \\
}

\date{}


\maketitle

\label{firstpage}
\begin{abstract}

We have observed four low-luminosity active galactic nuclei classified as Type 1 LINERs with the X-ray Telescope (XRT) and the UltraViolet-Optical Telescope (UVOT) onboard \sw, in an attempt  to clarify the main powering mechanism of this class of nearby sources.  Among our targets, we detect X-ray variability in NGC~3998 for the first time.  The light curves of this object reveal variations of up to 30\% amplitude in half a day, with no significant spectral variability on this time scale.  We also observe a decrease of $\sim$30\% over 9 days, with significant spectral softening.  Moreover, the X-ray flux is $\sim$40\% lower than observed in previous years.  Variability is detected in M~81 as well, at levels comparable to those reported previously: a flux increase in the hard X-rays (1-10 keV) of 30\% in $\sim$3 hours and variations by  up to a factor of 2 within a few years.    This X-ray behaviour is similar to that of higher-luminosity, Seyfert-type, objects.  Using previous  high-angular-resolution imaging data from the {\it Hubble Space Telescope} (HST), we evaluate the diffuse UV emission due to the host galaxy and isolate the nuclear flux in our UVOT observations.  All sources are detected in the UV band, at levels similar to those of the previous observations with HST.  The XRT (0.2--10\,keV) spectra are well described by single power-laws and the  UV-to-X-ray flux ratios are again consistent with those of Seyferts and radio-loud  AGNs of higher luminosity. The similarity in X-ray variability and broad-band energy distributions suggests the presence of similar accretion and radiation processes in low- and  high-luminosity AGNs.

\end{abstract}

\begin{keywords}
galaxies: active --- galaxies: nuclei --- ultraviolet: galaxies --- X-rays: galaxies

\noindent
Facility: {\it Swift}, {\it Chandra}

\end{keywords}

\section[]{Introduction\label{liners:intro}}

Low-luminosity active galactic nuclei (AGNs)  are a common phenomenon,
 with a large fraction of all
massive galaxies displaying some weak activity that is likely of non-stellar
origin (see, e.g., Ho 2008, and references therein).
Understanding the demographics and physics of these objects is a necessary
step towards the comprehension of super massive black hole activity in the local 
Universe and its past evolution. 
Low-Ionization Nuclear Emission Line Regions (LINERs), in particular,  
are a class of low-luminosity AGNs
defined on the basis of their optical spectral line ratios
\citep{Heckman1980:linerdef,Ho2008:llagnannrev}
and can be further divided into various subclasses according to their
different properties  (Chiaberge et al. 2005; Gonzalez-Martin et al. 2009). A fundamental question
about these sources is the origin of their optical spectrum and
multi-wavelength emission in general: it is not clear which fraction of the
powering source is non-stellar and, if it is due to accretion, what is the
regime of the accretion and the efficiency of the radiation conversion.     
It has been argued that the lack of X-ray variability  (e.g., Roberts et al. 1999; Komossa  et al. 1999; Georgantopoulos et al. 2002; Pellegrini et al. 2003; Ho 2008), the non detection of broad  Fe K$\alpha$ lines
down to stringent limits \citep{Ptak2004:ngc3998}, and the weakness or absence 
of the characteristic `big blue bump' in their optical/near-ultra-violet (UV) spectra,
traditionally observed in Seyferts 
\citep{Quataert1999:linersriaf,Chiaberge2006:ngc4565}, indicate that
 their engines may be intrinsically different from those of the more
 luminous AGNs,
 and could consist of radiatively
inefficient accretion flows (RIAFs).

However,  X-ray observations of most of these sources have been sparse
and not sensitive or not long  enough to detect significant variability, except in M~81, one of the closest and best studied {\it bona fide}  \citep[e.g.\ ][]{Kewley2006:linersratios} LINERs.  Furthermore, 
UV monitoring with the {\it Hubble Space Telescope} (HST) Advanced Camera for Surveys (ACS) 
of a sample of 17 LINERs has
revealed the presence of bright and variable UV nuclei \citep{Maoz2005}.  
By coupling these UV measurements with non simultaneous X-ray measurements
with {\it ASCA}, {\it Chandra}, and {\it XMM-Newton}, \citet{Maoz2007:lowlumagns} 
has shown that the 
UV-to-X-ray flux ratios in LINERs are similar to those of
much-more-luminous
Seyferts.
Thus, contrary to the common paradigm, rather than being qualitatively 
different from Seyferts, LINERs may instead be
 `scaled-down' analogues of Seyferts, with similar
 emission mechanisms operating in both classes.  

While the arguments of \citet{Maoz2007:lowlumagns} are supported by accurate photometry, 
the UV and X-ray data of his sample were not simultaneous, with measurements often separated by years.  
LINERs exhibit variability, which, albeit not of very large amplitude
(by factors of up to a few in the UV on time scales of years; \citealt{Maoz2005}),
may undermine the results derived
from non-simultaneous observations. 
In this paper, we approach this problem by obtaining, for the first
time, simultaneous X-ray and UV data, along with X-ray variability
information,
for a small sample of LINERs.   
The NASA satellite \sw\
is well suited to this task because it is a flexible and efficient
facility for long simultaneous and accurate UV and X-ray monitoring.
Results of this work have been presented in preliminary form in Romano et al. (2009).

Throughout this paper the uncertainties are given at  
90\% confidence levels  
for one interesting parameter (i.e., $\Delta \chi^2 =2.71$),  
unless otherwise stated.  
The spectral indices are parameterized as   
$F_{\nu} \propto \nu^{-\alpha}$, 
where $F_{\nu}$ (erg cm$^{-2}$ s$^{-1}$ Hz$^{-1}$) is the  
flux density as a function of frequency $\nu$; 
we also use $\Gamma = \alpha +1$ as the photon index,  
$N(E) \propto E^{-\Gamma}$ (ph cm$^{-2}$ s$^{-1}$\,keV$^{-1}$).  
This paper is organized as follows.  
In Section~2 we describe our sample,  observations, analysis, 
and results obtained with the two \sw\ instruments;  in Section~3 we construct the 
spectral energy distributions (SED);  and
in Section~4 we discuss our findings. 

\section[]{Sample, Observations, Data Analysis and Results \label{liners:data} }

Our sample consists of the brightest LINERs in 
the \citet{Maoz2007:lowlumagns} UV sample, 
but excluding M~87, because of its
prominent jet, which would dominate the UV emission at the \sw\
resolution. 
The sample includes M~81, NGC~3998, NGC~4203 and NGC~4579 (see
Table~\ref{liners:tab:specfits}), which are all type-1 LINERs (i.e., with detected broad 
H$\alpha$ components) according to the classification by 
\citet{Ho1997:linerspec}.  
These four objects form a rather homogeneous sample \citep[see ][]{Maoz2007:lowlumagns}:  
all have similar UV and X-ray luminosities ($10^{40}$ to $10^{41}$ erg~s$^{-1}$),
radio loudness parameters ($\sim100$), 
optical-to-X-ray indices $\alpha_{ox}$ ($\sim 1$), central black hole masses  
($10^7$ to $10^{8.4}$ M$_\odot$, three of which are  
measured directly from stellar or gas kinematics) 
and Eddington ratios ($-4.5$ to $-5.5$).  

The \sw\ archive\footnote{http://swift.gsfc.nasa.gov/docs/swift/archive/}  already contained data for two of these sources, M~81 and NGC~4203,  which 
had been observed as \sw\ fill-in targets, and which we retrieved.
M~81 was observed for 33.7\,ks and has an X-ray Telescope  
\citep[XRT, ][]{Burrows2005:XRTmnras} spectrum of
excellent quality.  
NGC~4203 had a good XRT spectrum, although somewhat under-exposed compared to
M~81 (5.3\,ks). 
We observed NGC~3998 and NGC~4579 with \sw\ for the first time as Targets of Opportunity (ToO) for 27.4 and 20.8\,ks, respectively.

\subsection[]{XRT \label{liners:data_xrt} }

Table~\ref{liners:tab:xrtobs} reports the log of the \sw/XRT 
observations used for this work. 
The XRT data were processed with standard procedures ({\sc xrtpipeline} 
v0.11.6), filtering and screening criteria by using {\sc  FTOOLS} in the
{\sc Heasoft} package (v.6.4). 
Given the low count rate of the sources during the respective observing 
campaigns ($<0.5$ counts s$^{-1}$),  
we only considered photon counting data (PC), and further selected XRT 
grades 0--12.  No pile-up correction was required. 
The source events were extracted in circular regions centered on the source, with 
radii depending on the source intensity (10--20 pixels, 1 pixel $\sim2\farcs37$),
while background events were extracted in source-free annular or circular regions,
depending on the field.  

Spectra were extracted for each XRT observation, as well as for the cumulative 
observing campaigns. 
Ancillary response files were generated with {\sc xrtmkarf},  
and account for different extraction regions, vignetting and 
Point-Spread Function (PSF) corrections. 
We used the v010 spectral redistribution matrices 
available in the Calibration Database maintained by HEASARC.
All spectra were rebinned with a minimum of 20 counts per energy bin 
to allow $\chi^2$ fitting within {\sc XSPEC} (v11.3.2).

We extracted XRT light curves from the same regions as for the spectra  
for source and background in the standard bands, 0.2--10\,keV (total), 
0.2--1\,keV (soft, S) and 1--10\,keV (hard, H). 
For our analysis, we considered several time bins ranging from 
120\,s to the typical orbit duration ($\la5800$\,s), and the minimum time bin was chosen  to ensure that each light curve point would have at least 
30 source-plus-background counts. 
The exposure requirements were that the bins be at least 50\,\% exposed.
In each case, the light curves were corrected for  
PSF losses, i.e., losses due to the extraction region geometry, 
and bad/hot pixels and columns falling within this region.

The host galaxy contribution in the XRT extraction region was evaluated 
using archival {\it Chandra} images of these LINERs obtained with the
longest exposures with ACIS-S. 
We evaluated the counts in the following regions: 
{\it i)}  R1, an annulus with outer radius equal to the XRT extraction radius for each object
and inner radius 4\arcsec, i.e., the radius including 99\% of the {\it Chandra} PSF;
{\it ii)} R2, a circular region of  4\arcsec radius;
{\it iii)} R3, an annular region of radii 4\arcsec and 4\farcs5.
We assumed that R1 only contained host galaxy photons, R2 only active nucleus photons,
while R3 was used to evaluate the galaxy counts in R2 assuming a constant host galaxy intensity profile.
The measured counts in R3 were scaled to the area of R2 and added to the counts in R1,
thus obtaining an estimate of the host galaxy counts in the whole XRT extraction region.
The host galaxy contribution to the X-ray emission of the four LINERS ranges 
from 6\,\% (NGC~3998) to 16\,\% (NGC~4203), and was therefore 
subsequently ignored.

\begin{figure*}
 	\includegraphics[angle=0,width=17cm,height=20cm]{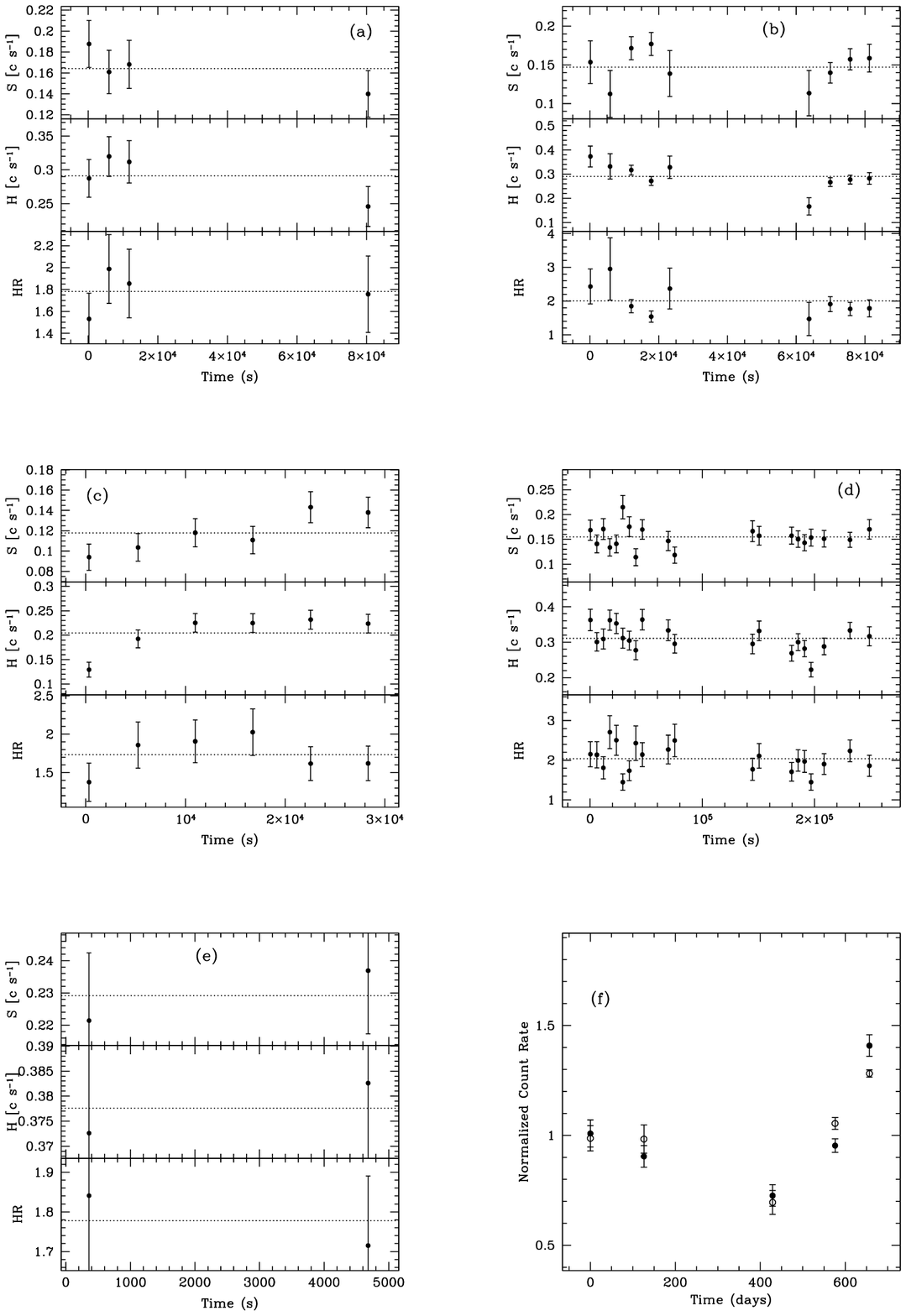}
	\caption{
	 \sw/XRT background-subtracted light curves (count rate in counts s$^{-1}$) and hardness ratios (i.e., 1-10 keV to 0.2-1 keV flux ratios)  of M~81.  The binning time interval corresponds to the orbit duration.  Panels {\it (a)} to {\it (e)} report  the  0.2--1\,keV (top), 1--10\,keV (middle) light curves and hardness ratio curves (bottom) of each pointing  (see Table 2).   Start times ($t = 0$) correspond to: {\it (a)}  2005 Apr 21.033 UT,  {\it (b)}  2005 Aug 25.056 UT, {\it (c)} 2006 Jun  24.004 UT, {\it (d)}  2006 Nov 18.045 UT, {\it (e)} 2007 Feb 6.906 UT.  
Panel  {\it (f)}  reports the  0.2--1\,keV (filled circles) and 1--10 keV (open circles) light curves of M~81 between 2005 and 2007.    Each point is the average of the flux measured during each pointing in that given band and each curve is normalized to its average (0.16 and 0.29 counts s$^{-1}$ for the 0.2--1\,keV and  1--10\,keV curves, respectively).
The time origin ($t=0$) corresponds to  2005 Apr 21.0 UT.   	}
	          \label{liners:fig:xlc_m81}
\end{figure*}


\begin{figure*}
 	\includegraphics[angle=0,width=17cm,height=20cm]{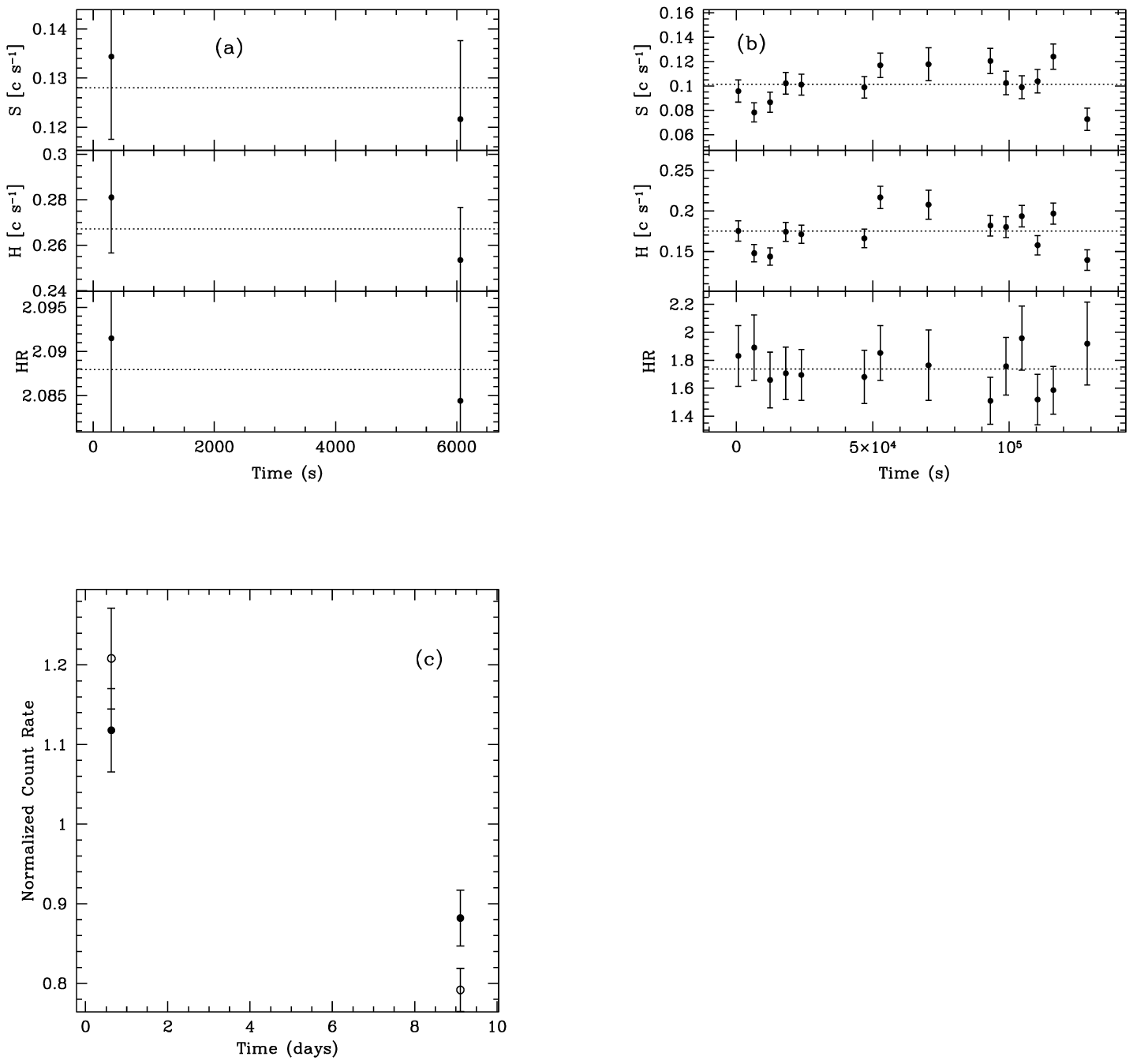}
	\caption{
\sw/XRT background-subtracted light curves (count rate in counts s$^{-1}$) and hardness ratios (i.e., 1-10 keV to 0.2-1 keV flux ratios)  of NGC~3998.  The binning time interval corresponds to the orbit duration.  Panels {\it (a)} and {\it (b)} report  the  0.2--1\,keV (top), 1--10\,keV (middle) light curves and hardness ratio curves (bottom) of each pointing  (see Table 2).  Start times ($t = 0$) correspond to: {\it (a)}  2007 Apr 20.628 UT,  {\it (b)}  2007 Apr 29.11 UT.  
Panel  {\it (c)}  reports the  0.2--1\,keV (filled circles) and 1--10 keV (open circles) light curves of NGC~3998 in 2007.    Each point is the average of the flux measured during each pointing in that given band and each curve is normalized to its average (0.11 and 0.22 counts s$^{-1}$ for the 0.2--1\,keV and  1--10\,keV curves, respectively).
The time origin ($t=0$) corresponds to  2007 Apr 20.0 UT.   	}
	          \label{liners:fig:xlc_n3998}
\end{figure*}



\begin{figure}
 	\includegraphics[angle=0,width=7cm,height=9cm]{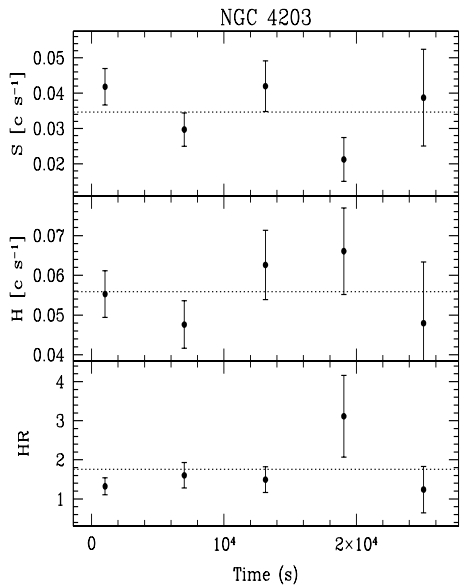}
	\caption{
	 \sw/XRT background-subtracted light curves (count rate in counts s$^{-1}$) and hardness ratios (i.e., 1-10 keV to 0.2-1 keV flux ratios) of NGC~4203: 0.2--1\,keV (top), 1--10\,keV (middle) light curves and hardness ratio curves (bottom).  The binning time interval corresponds to the orbit duration. The  time origin ($t = 0$) corresponds to  2005 Dec 25.004 UT.}
	          \label{liners:fig:xlc_n4203}
\end{figure}


\newpage

\begin{figure}
 	\includegraphics[angle=0,width=7cm,height=9cm]{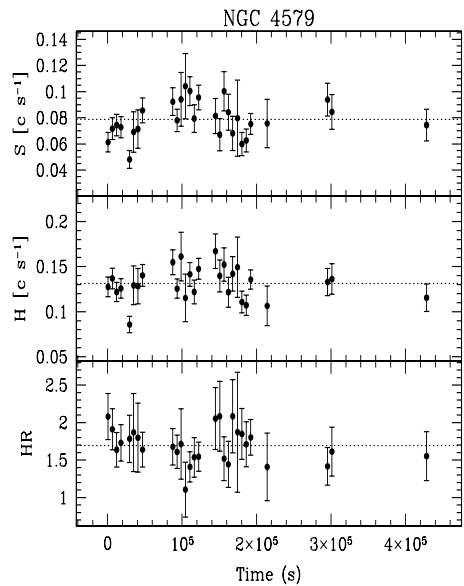}
	\caption{
	 \sw/XRT background-subtracted light curves (count rate in counts s$^{-1}$) and hardness ratios (i.e., 1-10 keV to 0.2-1 keV flux ratios) of NGC~4579 during the 4 \sw\ pointings at this source: 0.2--1\,keV (top), 1--10\,keV (middle) light curves and hardness ratio curves (bottom).  The binning time interval corresponds to the orbit duration. The  time origin ($t = 0$) corresponds to  2007 May 15.039 UT.}
	          \label{liners:fig:xlc_n4579}
\end{figure}


M~81 and NGC~3998 display a few very rapid (minute time scale) flux changes of 30 to 60\%. Although this exceeds the typical level of XRT photometric stability, which is about 10\% (as established on the basis of the flux stability of the standard source PSR~0540-69, Cusumano, priv. comm.),  we note  that these occur at low flux levels, so that we cannot assess their authenticity.   Therefore,  we have considered only fluxes integrated over time scales no shorter than one orbit for further variability analysis.

Since the \sw\ satellite must re-settle on the target at every orbit,
the target's location within the XRT field of view may change from orbit to orbit, and occasionally the source flux is greatly underestimated because of the vicinity of a detector dead column that affects the PSF to various degrees.  We have systematically checked our observations for this effect  and have corrected for it (see Grupe et al. 2007).  However, because of the low flux level of our sources,  in many cases the PSF reconstruction is not  fully satisfactory and the corresponding flux points have been excluded from our variability analysis.  

The orbit-averaged light curves of M~81 and NGC~3998 indicate
variability.  In Figure 1c, 1d and 2b we show the flux and hardness-ratio time series of M~81 and NGC~3998 during the pointings when significant variations were observed.
The M~81 hard X-ray light curves in June (Fig. 1c, middle panel) and  November (Fig. 1d, middle panel)  2006 have $\chi^2 \simeq 30$ (for 5 degrees of freedom, d.o.f.) and $\chi^2 \simeq 41$  (for 19 d.o.f.) with respect to the average flux, respectively, corresponding to probabilities of constancy of $10^{-3}$ or less.  The variations in hard X-rays have a maximum amplitude of  30\% in a time scale of 3 to 12 hours.   The variability in soft X-rays is less pronounced, but well correlated with the hard X-rays, so that the hardness ratio  is not significantly variable on these time scales.    
The soft and hard X-ray light curves of NGC~3998 on 29 April 2007 (Fig. 2b)  are well correlated and  have  $\chi^2 \simeq 35$  (13 d.o.f.) and $\chi^2 \simeq 44$ (13 d.o.f.)  with respect to the average flux, respectively, corresponding to constancy  probabilities of $10^{-3}$ or less. The  largest variability amplitude of NGC~3998 is about 30\% in both X-ray energy ranges, on a time scale of about 12 hours.    Neither NGC~4203 (Fig. 3) nor NGC~4579 (Fig. 4) vary significantly in flux or hardness ratio on the orbit or longer time scale.  

We have evaluated the variability of M~81 and NGC~3998 during the
individual \sw\ pointings also through the excess variance parameter
$\sigma_{rms}^2$,  defined as in  Nandra et al. (1997) and Turner et
al. (1999; see also Ptak et al. 1998).  This never exceeds the value
of 0.02 for both sources and for both energy ranges, and therefore it
is almost one order of magnitude smaller than would be expected -- 
based on the X-ray luminosities of these two LINERs --  from the inverse proportionality relation between X-ray luminosity  and  variability established for Seyfert galaxies by Nandra et al. (1997) and Turner et al. (1999).\footnote{Note that the durations of our XRT observations are similar to the monitoring times adopted by those authors, making the comparison of the $\sigma_{rms}^2$ parameters  meaningful.}

For M~81 and NGC~3998, which were observed on more than one epoch, 
with intervals larger than 1 day, we have evaluated the average flux at each epoch. 
The associated errors   are the root mean squares of the averages, i.e. the standard deviations of the data points within the epoch, divided by the square root
of their number.   The corresponding light curves are shown in Figures 1f and 2c for M~81 and NGC~3998, respectively. The long term variations  are conspicuous.  The historical flux of M~81  decreases between 
2005 April and 2006 June almost achromatically and then increases by 2007 February by a factor of 2 in the
soft X-rays and slightly less in hard X-rays.  The variability on year time scales is well correlated in the
two bands, with no significant variations of the hardness ratio.     The 1--10\,keV flux  of NGC~3998 decreases by 30\% in the nine days separating our two 
observations,  while the 0.2--1\,keV flux decreases more slowly, implying a softening of the spectrum.  Indeed,  the hardness ratio (i.e., the 1-10 keV  to  0.2-1 keV flux ratio) decreases significantly between the 2 epochs, varying from HR = $2.088 \pm 0.004$ to HR = $1.75 \pm 0.04$.    However, a spectral fit with a single power-law of each observation of this object does not show variability in the spectral parameters
within the 90\% confidence level.  

The mean XRT spectrum for each object was fit with a simple absorbed power-law
model with free absorption and photon index.
Figure~5 shows the mean spectrum for each object
in the sample, along with its best-fit model, while the fit results are reported
in Table~\ref{liners:tab:specfits}. The spectra show no significant
absorption features superimposed on the power-law continuum; however,
M~81 shows hints of an Fe K$\alpha$ emission line (6.4\,keV, fixed), with an equivalent width
EW $=640_{-410}^{+650}$\,eV. The F-test probability for this feature is
$8.515\times 10^{-3}$, which corresponds to a 2.6-$\sigma$ detection.
NGC~4579 also shows a marginal (F-test probability of $3.207\times10^{-2}$ or 2.1-$\sigma$)
Fe K$\alpha$ line at $6.4_{-0.4}^{+0.2}$\,keV, with EW $=550_{-310}^{+5100}$\,eV.
The fluxes of these lines are
$(10^{+10}_{-4})\times 10^{-14}$ and $(5^{+32}_{-2})\times 10^{-14}$
erg\,cm$^{-2}$\,s$^{-1}$, respectively.
The equivalent widths  are, within 
their large errors, broadly consistent  with the values reported by 
\citet{Pellegrini2000a:m81bsax}  and \citet{Terashima2002:ASCAliners}.

Although the uncertainties on the fitted hydrogen column densities
(Table~\ref{liners:tab:specfits}, column\ 4)
are not small, all sources -- except NGC~4203, whose spectrum  has a  poor signal-to-noise ratio  --  show significant evidence for absorption in excess of
the Galactic one (also reported in Table~\ref{liners:tab:specfits}).

\begin{figure*}                 
	\vspace{-5truecm}
\includegraphics[angle=0,width=18cm]{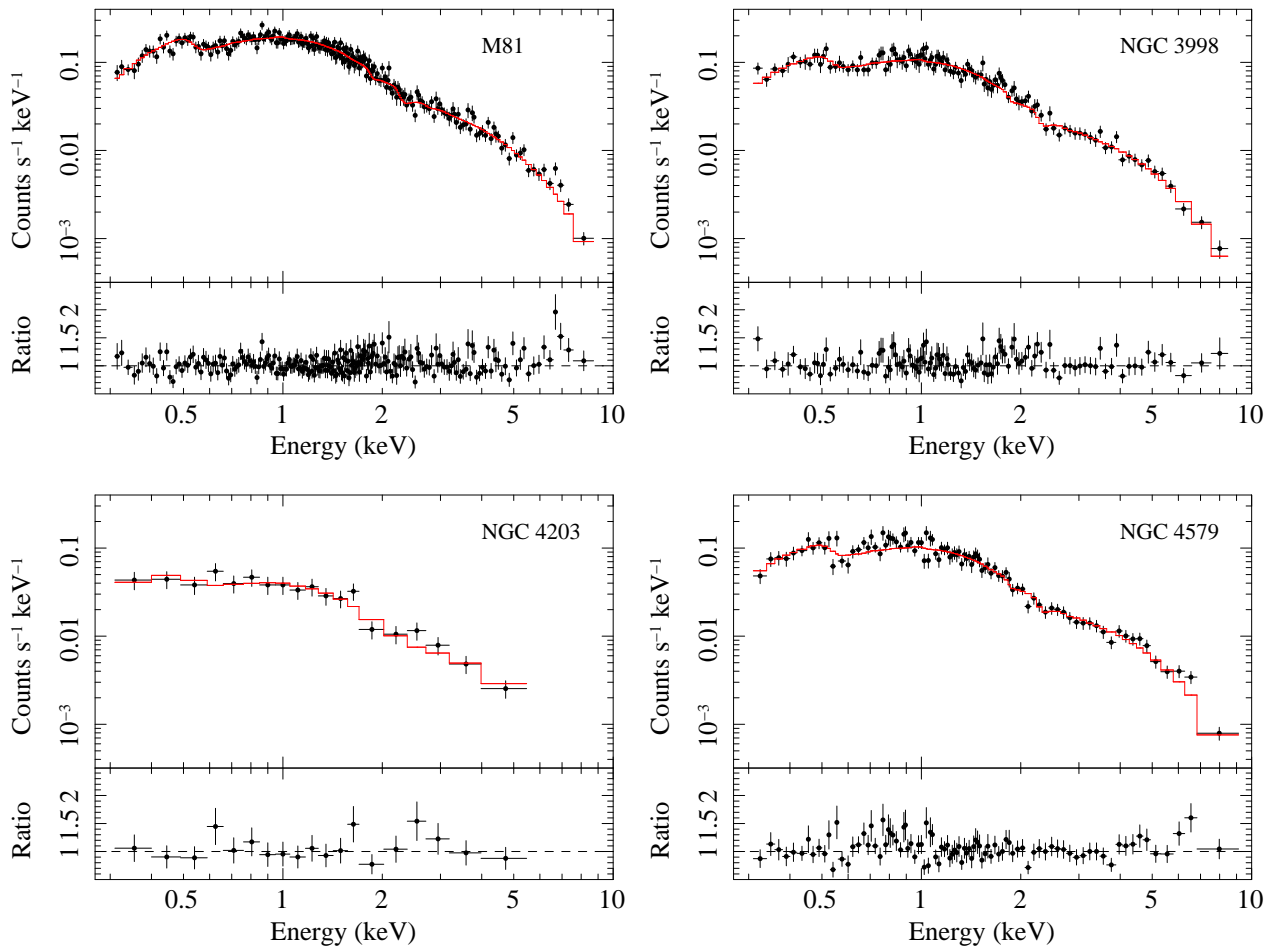} 
	\vspace{-7truecm}
\caption{\sw/XRT spectra of the sample. 
		The stepped curves represent the single absorbed power-laws 
		which best fit the spectra (see  Table~\ref{liners:tab:specfits} for 	
		spectral parameters).
} 
\label{liners:fig:xrtspectra}
\end{figure*}

\subsection[]{UVOT \label{liners:data_uvot} }

The \sw\ UltraViolet-Optical Telescope \citep[UVOT, ][]{Roming2005:UVOTmnras} 
observed the four targets with the filters $u$ (3465\,\AA)
$uvw1$ (2600\,\AA), $uvm2$ (2246\,\AA), $uvw2$ (1928\,\AA) 
simultaneously with the XRT.   The filter choice was driven by the objective of maximizing 
the nuclear signal, which is dominant in the UV,  while 
minimizing the stellar emission from the bulge populations of the target galaxies.
For each object, after verifying that the UVOT counts  show no significant variability, 
we have coadded the images.

The data analysis was performed using the {\sc uvotsource} 
task included in the latest {\sc Heasoft} software. 
This task normally calculates the magnitude by means of aperture 
photometry within a circular region of $5^{\prime\prime}$-radius, 
which includes more than 99\,\% of the signal of a point source,
and applies specific 
corrections due to the detector characteristics.  Since the adoption
of this standard aperture often 
resulted in over-subtraction of the host-galaxy background, we used
instead an
aperture of $2^{\prime\prime}$ radius, and applied the proper
correction factor between a $2^{\prime\prime}$-radius aperture 
to a $5^{\prime\prime}$-radius aperture
(see below).
Because of its proximity (3.6\,Mpc), 
the host galaxy of M~81 occupies most of the field-of-view of UVOT, 
making it difficult to estimate the sky background to be subtracted 
from the nucleus and the galaxy light.
The background was therefore computed from a $20\farcs7$-radius circular region
separated by $6\farcm5$ from the nucleus. 
For uniformity, the remaining three LINERs, albeit more distant, were treated in the same way.

Using archival HST images of 
these LINERs at 2500\,\AA\ and 3300\,\AA\  obtained with the High Resolution Channel (HRC) of the  HST
ACS \citep{Maoz2005}, we have evaluated the
nuclear-point-source-subtracted host 
galaxy contribution for each object
in a $2\,\arcsec$-radius circular area centred on the nucleus,
and have  interpolated or 
extrapolated the UV host galaxy fluxes to the central wavelengths of the UVOT $u$-band 
and UV filters, in order to subtract them from the observed fluxes.

While the aperture-summed galaxy light is clearly detected in the HST
images, the surface brightness is too low and the images are too oversampled
to determine the surface brightness profiles of the galaxies on the
scales of the apertures we use here. Different galaxy profiles will require
different corrections for PSF losses outside the aperture. For the
galaxy light subtraction, we have therefore considered two extreme
cases: one in which the galaxy light is highly centrally concentrated,
and one in which it has constant surface brightness. In the former
scenario, the adoption of a $2\,\arcsec$-radius for the UVOT photometry
results in a galaxy light loss similar to that suffered by the nuclear
light, and therefore we first corrected the observed
$2\,\arcsec$-radius flux to a 
$5\,\arcsec$-radius aperture, and then subtracted the galaxy flux
evaluated from the HST images.  In the latter case, where the light
profile is flat, the galaxy light suffers no net PSF losses
in the UVOT photometry, and we therefore first subtracted
the HST galaxy flux from the total observed UVOT flux and then applied 
the $5^{\prime\prime}$-radius aperture correction to the remainder,
which represents the point-source nuclear flux.

The host galaxy contribution is dominant in the $u$ band (it is generally comparable to 
the total flux observed by UVOT) and decreases
with decreasing wavelength to the $uvw2$ filter 
(1928\,\AA) where  it contributes, at most, 50\,\% of the observed flux.   The first galaxy-subtraction method described above
predictably results in systematically larger nuclear fluxes
than obtained with the second method. As also expected, the
differences between the results of the two methods decrease as one
goes to  shorter wavelengths, where the galaxy light becomes less dominant.

All UVOT magnitudes have been converted into fluxes using the 
latest in-flight flux calibration factors and zero-points \citep{Poole2008:uvotconverfac}.
The average observed fluxes  with their statistical errors are 
reported in Table \ref{liners:tab:uvotobs}, where we also list the host galaxy fluxes 
measured in the HST images and converted to the wavelengths of the UVOT filters,  and the final nuclear fluxes obtained with the two subtraction methods described above.

\section[]{SED Construction \label{liners:seds} }

We combined the XRT spectra, corrected for the total hydrogen absorption 
(Table~\ref{liners:tab:specfits}, col.\ 4), 
and the UVOT fluxes, corrected for their host galaxy backgrounds and 
dereddened using the $A_B$ extinction values compiled by
\cite{Maoz2007:lowlumagns} from Schlegel, Finkbeiner, \& Davis (1998),
and adopting the Galactic extinction curve of Cardelli, Clayton and Mathis (1989).   
These UV-to-X-ray spectral energy distributions (SED)  are shown
in Fig.~6.
The errors associated with the UVOT nuclear fluxes are the sum in quadrature 
of the uncertainties associated with the observed UVOT fluxes [including the statistical 
error reported in 
Table \ref{liners:tab:uvotobs} and the systematic errors given in 
\citet{Poole2008:uvotconverfac}]  and the uncertainties of the
galaxy HST measurements (Table \ref{liners:tab:uvotobs}),  increased
by  5\,\% -- summed 
in quadrature --  to take  into account the uncertainty in the background subtraction 
on the HST images and the interpolation/extrapolation errors.
Whenever the galaxy flux  evaluated from the HST images exceeded the
total flux observed by UVOT,   
we have used the total uncertainty to compute  a 3-$\sigma$ upper
limit, which is indicated in the figure.

For comparison, we have plotted also the previously measured UV-to-X-ray 
measurements [see Fig.~1 in \citet{Maoz2007:lowlumagns}] and the mean SEDs of 
radio-loud and radio-quiet 
quasars from \citet{Elvis1994:RLRQmeantem}, normalized to go through the geometric mean of the two nuclear fluxes  
at 1928\,\AA\ obtained with the two galaxy subtraction methods described above.    Note that the 
fluxes in this band suffer the least galaxy contamination, 
and therefore represent a solid 
lower limit to the intensity of the UV component.   

\begin{figure*}
	\vspace{-5truecm}
 	\includegraphics[angle=0,width=20cm]{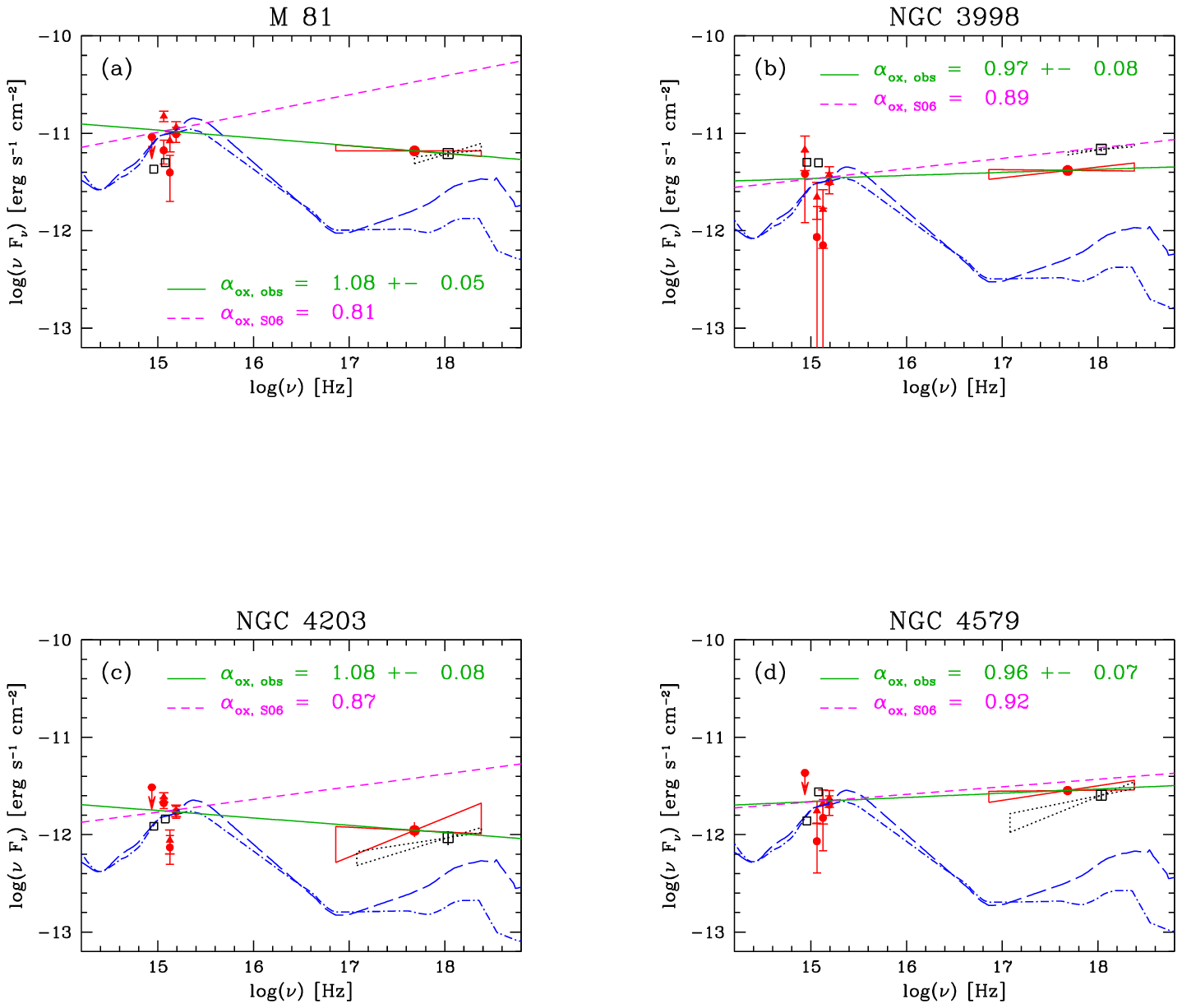}
	\vspace{-6truecm}
	\caption{Spectral energy distributions of our sample based on the simultaneous 
\sw/XRT and UVOT data from the present campaign (filled symbols) and from 
previous, non-simultaneous, {\it HST} UV and {\it Chandra}, {\it XMM-Newton} or {\it ASCA} 
X-ray observations [empty squares; \citet{Ho2001:cxoliners,Terashima2002:ASCAliners,Ptak2004:ngc3998,LaParola2004:CXOm81,Maoz2005,Cappi2006:XMMaa,Maoz2007:lowlumagns}].
The optical-UV data, corrected for Galactic reddening, result from
the observed UVOT fluxes, minus the HST host galaxy fluxes 
assuming that the host galaxy profile is strongly centrally concentrated (filled triangles) or
constant with radius within the UVOT PSF (filled circles).   When the
latter method yields a negative flux (see  Table~3), we replace both nuclear fluxes with a 3-$\sigma$ upper limit.  
The X-ray data are corrected for both Galactic and 
intrinsic absorption.    The power laws fitting the  XRT (solid bow-tie) and archival (dotted bow-tie)  X-ray spectra are reported with their 90\% confidence 
boundaries.  Similarly to \citet{Maoz2007:lowlumagns}, we have superimposed on the data the 
mean spectra of radio-loud (long-dashed curve) and radio-quiet (dot-dashed curve) quasars 
from \citet{Elvis1994:RLRQmeantem}, and we have normalized them to
the geometric mean of the 1928\,\AA\  nuclear fluxes obtained with the
two methods of galaxy subtraction.  We also show lines
that connect the 2500~\AA\  and 2 keV  fluxes  (solid)
and the lines predicted by the S06 relation
between $\alpha_{ox}$ and the monochromatic UV luminosities
(short dash).     }
          \label{liners:fig:4sed}
\end{figure*}

\section[]{Discussion\label{liners:discussion}}

We have observed four  bright LINERs with XRT and UVOT onboard \sw, 
to study their variability and their simultaneous UV-to-X-ray ratios, in
an attempt 
to probe their emission mechanisms.   In particular, 
high-luminosity AGNs -- Seyferts and QSOs -- are generally thought to 
be powered by geometrically thin and 
optically thick accretion disks.
Our new data can bring new insight to the question of whether
low-luminosity AGNs -- specifically LINERs -- are simply 
lower-power analogues of Seyferts and QSOs, or  are instead 
powered by some fundamentally distinct process, such as a RIAF.    

Among our results is the first detection of X-ray variability
in NGC~3998, in a range of time scales from a few hours to days, with a
$\sim$30\% amplitude.   This is in contrast to earlier reports of no
X-ray variability during  BeppoSAX and XMM-Newton observations
\citep{Pellegrini2000b:ngc3998bsax,Ptak2004:ngc3998}.   The reason for
this difference may be related to the larger sensitivity  of XRT with
respect to BeppoSAX, and to XRT's higher scheduling flexibility, which
allows for longer monitoring periods than possible with XMM-Newton.  
\citet{Pellegrini2000b:ngc3998bsax} have cited
the absence of variability in NGC~3998 as evidence that the
accretion is advection-dominated in this source. Other workers 
(Ptak et al. 1998; Ho 2008) have, in general, invoked
non-variability in X-rays as an indicator of a RIAF mode.
Our finding of significant short-term X-ray flux variations in a low-luminosity
LINER points, instead, to a similarity with whatever accretion mode
is occurring in higher luminosity AGNs.    However, the excess variance
parameter of the NGC~3998 XRT light curves does not obey
the inverse relation between X-ray luminosity and variability
amplitude determined for high-luminosity AGNs (Nandra et al.
1997; Turner et al. 1999) when this is extrapolated to the X-ray luminosity of NGC~3998 ($1.1 \times 10^{40}$ erg~s$^{-1}$ in  2-10 keV). 
Instead, the variability amplitude is similar to that typical of 
Seyferts of luminosity $10^{43}-10^{44}~{\rm erg~s}^{-1}$ (2-10 keV).
The range of validity of the relation may thus not extend to the  
low X-ray luminosities of the objects studied here.

A second LINER in our sample, M~81, has already been
 shown to be X-ray variable on many time
scales, from hours to years (Pellegrini et al. 2000a; La Parola et
 al. 2004; Page et al. 2004; Markoff et al. 2008; Iyomoto \& Makishima
 2001; Young et al. 2007). Our findings confirm this behaviour and set
this object, along with NGC~3998, as another X-ray-variable
 low-luminosity AGN that is similar in this respect to high-luminosity AGNs.  
It has been suggested that the mechanisms active in this object are
 similar
to those that cause the hard states in X-ray binaries 
\citep{Markoff2008:m81bhscal}.  Our observation of spectral softening accompanying source brightening  on a time scale of years is indeed reminiscent of the behavior exhibited, on a shorter time scale, by Galactic sources. 

Our XRT timing results on NGC~4203 and NGC~4579 are inconclusive.
NGC~4203 exhibits no significant X-ray variability on any time scale, despite being, among the sources in 
\citet{Maoz2007:lowlumagns}, the one that varied with the largest amplitude in the UV. 
One reason may be that our monitoring period was limited, 
and hence we did not catch the stochastically occurring variations.
Alternatively, for this object the errors associated with orbit-averaged points are
somewhat larger than those of M~81 and NGC~3998, so that inter-orbit
variations  of amplitude similar to the inter-orbit variations in M~81
and NGC~3998 would not be detected at high significance. 
A prominent, variable Fe K$\alpha$ emission line has been
detected by  {\it ASCA} and {\it XMM-Newton} in NGC~4579, and has given rise to various
interpretations as to its origin from a standard accretion disk or a
truncated  disk 
\citep{Dewangan2004:ngc4579xmmfeline,Terashima1998:ngc4579ascafeline,Terashima2000:ngc4579felinevar,Terashima2002:ASCAliners}.
Our weak XRT detection  of the feature 
does not add significantly to this debate.

Our sources are all well detected in at least one UV filter
(Fig.~6).   The  UV-to-X-ray SEDs compare rather well with those reported by \citet{Maoz2007:lowlumagns}
and obtained with non simultaneous UV and X-ray data.
The X-ray spectrum of the LINERs always lies about a factor of 3 to 10 
above the normalized 
radio-loud quasar template, depending on the object and on the
wavelength.   The UV flux level
has not changed dramatically in any of the sources 
(Fig.~6),  although these are among the most UV-variable sources in the 
\citet{Maoz2007:lowlumagns} sample.  We note, however, that the UVOT fluxes are more uncertain
than the HST measurements, owing to the galaxy subtraction and to the lower angular resolution 
and photometric accuracy of UVOT. The XRT fluxes and
spectra are also similar to the older ones, except for NGC~3998, which XRT
has detected in a state lower by $\sim$40\%   than observed  in 1999 and 2001 by BeppoSAX and XMM-Newton, respectively (Ptak et al. 2004, see also Fig.~6b).   

We have evaluated the Eddington ratios of our LINERs using the bolometric
luminosities and the  central black hole mass estimates  compiled in 
\citet{Maoz2007:lowlumagns}.  The bolometric luminosities were obtained from the
normalized templates of radio-loud AGNs \citep{Elvis1994:RLRQmeantem} from 3465\,\AA\
to 0.2\,keV and the observed X-ray spectrum from 0.2 to 10\,keV.  In fact,
the template may be a more reliable description of the spectrum at UV
wavelengths (considering that some of our UV points are upper limits),
while at X-rays the template always underestimates the real flux  (see
Fig.~6).   
In the radio-loud AGN spectrum, the UV-to-X-ray flux
dominates the SED, representing more than half of the total emission.  
Moreover, considering that the {\it intrinsic} AGN spectra probably lack
the conspicuous {\it observed} bump centred at $1\,\mu$m, due to
reprocessed radiation at shorter wavelengths \citep{Marconi2004:locsmbhagn}, our
computed bolometric luminosities may adequately represent  the 
intrinsic luminosities after all.  
The Eddington ratios, reported in Table~\ref{liners:tab:specfits},  
are about 0.5\,dex higher than
those computed by \citet{Maoz2007:lowlumagns}, who only included the UV luminosity, but
are still low, as typical for LINERs.   

By normalizing the AGN templates to the UV emission, we note that the
precisely determined XRT spectral slopes follow the average radio-loud AGN
shape, although, as noted, the spectral normalizations exceed the prediction by 0.5 to 1
order of magnitude (Fig.~6).  Strateva et al. (2005) and Steffen et al. (2006, 
hereafter S06) have studied the dependence of the optical-to-X-ray colour index, 
$\alpha_{ox}$, of AGNs (defined  -- using our notation for the spectral indices, see last paragraph of Section 1 -- as 
$\alpha_{ox}$ = --0.3838 log($l_{2 keV}/l_{2500 {\rm \AA}}$), 
where $l_{2 keV}$ and 
$l_{2500 {\rm \AA}}$ are the rest-frame X-ray and UV  monochromatic
luminosities  in erg~s$^{-1}$~Hz$^{-1}$,  respectively) on the monochromatic rest-frame UV
luminosity, and have found an empirical linear relation between those
quantities (the parameters of the relation in the two works
coincide, within the errors; we have used that of S06, which is based
on a larger sample).  We have extrapolated  $\alpha_{ox}$ in the S06  relation, 
$\alpha_{ox}$ = 0.137 log ($l_{2500 {\rm \AA}}$) -- 2.638
(note that we have changed the sign of their Eq. 2, to be coherent with our spectral index notation),  by two orders of magnitude to low 
UV luminosities, corresponding
to the observed monochromatic UV luminosities of the LINERs in our sample
(extinction-corrected, see Table 1, col. 9).  
This extrapolation is plotted as a
short-dashed line in Fig.~6.  
It can be compared to a power law passing through our
{\it observed}  luminosities at 2500 \AA\
and 2 keV, shown in Fig.~6 as
a solid line, with errors omitted, for clarity.  

In NGC~3998 and NGC~4579 the observed values of $\alpha_{ox}$   coincide, to
 within the errors, with the extrapolation to low luminosities of the S06 relation.   In M~81  and NGC~4203, the less UV-luminous sources in our sample (see
 Table~1),   $\alpha_{ox,obs}$ has an intermediate value, similar to those of Seyferts,  and
 the extrapolated $\alpha_{ox}$ is significantly flatter than the
 observed one and predicts X-ray luminosities  that are 4-5 times
 higher  than observed. As already noted by S06 based on their data alone,
and further pointed out by Maoz (2007) based on his non simultaneous
 UV and X-ray measurements,  
 the S06 relation probably flattens at luminosities below 
$\sim 10^{26}$ erg~s$^{-1}$~Hz$^{-1}$. Overall, our present simultaneous UV
 and X-ray photometry strengthens the case for a limiting value of
$\alpha_{ox}$ below some critical luminosity typical of Seyferts. 


In summary, for the LINERs under study, the significant ultraviolet emission, 
the X-ray intraday variability, and
the X-ray and multiwavelength spectral similarity  with radio-loud
AGNs all point to continuity and similarity with AGNs of higher luminosity.
In view of this, it is not clear that distinct accretion and radiation
mechanisms are required in the different luminosity regimes.
Further  \sw/XRT and UVOT monitoring, or more optimally, new HST high angular resolution
UV observations of larger samples, accompanied with XRT simultaneous coverage,
could shed more light on these questions.

We note, finally, that 
we cannot exclude that, in these sources, the emission at X-ray and/or
UV bands is partially due to a weak jet,  some evidence of which has been reported for all of our four sources, either  directly (for M~81, based on resolved radio structure detection and radio polarization)
or indirectly (for NGC~3998, NGC~4203, NGC~4579, based on a flat radio
spectrum, a compact and variable radio core, and high brightness temperature) 
from high resolution radio imaging (Bietenholz et al. 2000; Bower et al. 2002; Filho et al. 2002; Anderson et al. 2004; Bietenholz et al. 2004; 
Ros \& Perez Torres 2008).
In the case of NGC~3998, a jet has been proposed by Ptak et al. (2004) as possibly responsible for the multiwavelength emission.  Jetted high energy 
radiation in this LINER  would be  compatible  both with the  rapid flux variations  and with
the frequency-dependent variability amplitude that we observe over 9
days (Fig.~2c), which is typical of non-thermal sources
\citep[e.g. ][]{Ulrich:araablazars}, and which  is dissimilar to
the X-ray variability observed in
Seyferts [usually of larger amplitude at softer energies, 
e.g., \citet{Matsuoka1990:ginga,Nandra1990:ginga,Edelson 2000:ngc3516,Turner:Ark564,Uttley2005,Terashima2009:suzaku}]
and in weak-line radio-galaxies [e.g.,
  \citet{Gliozzi2008:llagnxmm}].   However, it would be premature to draw a conclusion on the exact
nature of the X-ray emitting process from the detection of spectral softening accompanying flux decrease between only two epochs in this individual LINER.  More work on the correlated radio and X-ray variability in LINERs -- in both the 
observational and the theoretical directions -- 
is necessary to
assess the role of a jet in the multiwavelength emission (see e.g.  Brenneman et al. 2009).


\section*{Acknowledgments}
We thank N.\ Gehrels for approving this set of ToOs and the \sw\ team, in particular the duty scientists and science planners, for making these observations possible.  We are grateful to S.\ Immler and T.\ Belloni for kindly sharing their data on M~81 and NGC~4203, G.~Cusumano, D.~Grupe,  S.\ Holland, and J.~Nousek for advice on the \sw\ instruments, and to C.~Guidorzi for helpful discussion.   We want to thank the anonymous referee for constructive comments that helped improving our presentation. PR thanks INAF-IASFMi,  and DM thanks INAF -- Osservatorio Astrofisico di Arcetri, where parts of this work were carried out, for their kind hospitality.  This work was supported by grants ASI-INAF I/023/05/0 and ASI I/088/06/0.  DM thanks the DFG for support via German-Israeli Project Cooperation grant STE1869/1-1.GE625/15-1.  This research has made use of NASA's Astrophysics Data System Bibliographic Services,  as well as the NASA/IPAC Extragalactic Database (NED), which is operated  by the Jet Propulsion Laboratory, California Institute of Technology, under contract with  the National Aeronautics and Space Administration.

\setcounter{table}{0}
\begin{table*} 	
  \begin{center} 	
 \caption{Target  parameters.  } 	
 \label{liners:tab:specfits} 	
 \begin{tabular}{lrrrrrrrrr} 
 \hline 
 \hline 
 \noalign{\smallskip} 
 Name	& Distance & $N^{\rm G,a}_{\rm H}$	&$N^b_{\rm H}$&		$\Gamma^c$& 		$\chi^2_{\rm red}/$dof	&	$F^{\mathrm{d}}$ & $F^{\mathrm{d}}$	 & $L_{UV}^e$  & $L/L_{Edd}^f$ \\
 	& (Mpc) & $(10^{20}$ cm$^{-2})$	&$(10^{20}$ cm$^{-2})$ 	&		   & &(0.2--1\,keV)	& (1--10\,keV)	& & \\
  \noalign{\smallskip} 
 \hline 
 \noalign{\smallskip} 
M~81      & 3.6 & 5.55	& $10.45_{-0.89}^{+0.93}$ & $2.04 \pm 0.04$ & 0.966/290 & $11.2 \pm 0.1$ & $14.9 \pm 0.4$ & 1.4 & $8.8 \times 10^{-6}$  \\ 

NGC~3998 & 13.1 & 1.01	&$ 6.85_{-1.20}^{+1.26}$ &$1.95 \pm 0.06$ & 0.948/188 &$6.2_{-0.8}^{+0.1}$  &$9.8_{-0.4}^{+0.5}$  & 6.0 & $1.1 \times 10^{-5}$  \\ 
NGC~4203 & 15.1 & 1.11	&$ 2.74_{-2.74}^{+4.51}$ &$1.81_{-0.21}^{+0.24}$ & 0.699/16  &$1.4_{-0.4}^{+0.2}$  &$2.8_{-0.4}^{+0.6}$  & 4.0 & $1.1  \times 10^{-4}$  \\
NGC~4579 & 21 & 2.97	&$ 7.01_{-1.34}^{+1.43}$ &$1.92 \pm 0.07$ & 0.972/145 &$4.1 \pm 0.1$  &$6.8 \pm 0.4$  & 9.8 & $7.4 \times 10^{-5}$  \\ 
 \noalign{\smallskip}
  \hline
  \end{tabular}
  \end{center}
  \begin{list}{}{} 
  \item[$^{\mathrm{a}}$]  Hydrogen column densities derived from \citet{Kalberla2005:nhgalsur} and consistent with the $A_B$ extinction 
values reported in \citep{Maoz2007:lowlumagns}, using a typical Milky Way gas-to-dust ratio, $5 \times 10^{21}$ cm$^{-2}$ mag$^{-1}$.     \item[$^{\mathrm{b}}$] Hydrogen column densities from the XRT spectral fits.
        \item[$^{\mathrm{c}}$]  Photon index, $f_E \propto E^{-\Gamma}$. 
        \item[$^{\mathrm{d}}$]  Unabsorbed flux in units of $10^{-12}$ erg cm$^{-2}$ s$^{-1}$.            
      \item[$^{\mathrm{e}}$]  Unabsorbed monochromatic luminosity at 2500 \AA, in  units of 
      $10^{25}$ erg~s$^{-1}$~Hz$^{-1}$.            
        \item[$^{\mathrm{f}}$]  Eddington ratio based on the observed SEDs. 
  \end{list} 
  \end{table*}

 \setcounter{table}{1}
\begin{table*}
 \begin{center} 	
 \caption{XRT observation log.} 	
 \label{liners:tab:xrtobs} 	
 \begin{tabular}{lrrrr} 
 \hline 
 \hline 
 \noalign{\smallskip} 
Name 		& Sequence$^{\mathrm{a}}$  & Start time  (UT)     	& 	End time   (UT)         &  Exposure  \\
     		&              & (yyyy-mm-dd hh:mm:ss)  & (yyyy-mm-dd hh:mm:ss)        	&(s)  \\
 \noalign{\smallskip} 
 \hline 
 \noalign{\smallskip} 
M~81 		& 00035059001	&2005-04-21 00:47:46	&	2005-04-21 23:11:33	&	1563	\\
		& 00035059002	&2005-08-25 01:20:04	&	2005-08-25 23:59:59	&	5022	\\ 
		& 00035059003	&2006-06-24 00:05:06	&	2006-06-24 08:02:57	&	4140	\\
		& 00035059004	&2006-11-18 01:05:11	&	2006-11-20 23:54:55	&	20986	\\
		& 00259527001	&2007-02-06 21:44:17	&	2007-02-06 23:13:04	&	1973	\\
NGC~3998	& 00030916001	&2007-04-20 15:03:58	&	2007-04-20 16:49:57	&	1198	\\
		& 00030916002	&2007-04-29 02:39:06	&	2007-04-30 14:32:27	&	26216	\\
NGC~4203	& 00035477001	&2005-12-25 00:05:03	&	2005-12-25 07:04:59	&	5261	\\
NGC~4579	& 00030939001   &2007-05-15 00:56:39    &       2007-05-15 14:07:01     &       7712      \\
		& 00030939002   &2007-05-16 01:08:52    &       2007-05-16 23:39:56     &       7870      \\
		& 00030939003   &2007-05-17 01:18:44    &       2007-05-17 12:30:58     &       3490      \\
		& 00030939004   &2007-05-18 10:53:03    &       2007-05-19 23:59:23     &       1743      \\
  \noalign{\smallskip}
  \hline
  \end{tabular}
  \end{center}
  \begin{list}{}{} 
  \item[$^{\mathrm{a}}$] We only considered data collected in PC observing mode.
  \end{list} 
  \end{table*}

\setcounter{table}{2}
\begin{table*} 	
  \begin{center} 	
 \caption{\sw/UVOT Observations.} 	
 \label{liners:tab:uvotobs}
 \begin{tabular}{lrrrrrr} 
 \hline 
 \hline 
 \noalign{\smallskip} 
Name    & Band	& Wavelength(\AA)          & Observed flux$^{\mathrm{a,b}}$ 	 	 & Galaxy flux$^{\mathrm{a,c}}$	&
Nuclear flux(1)$^{\mathrm{a,d,f}}$ & Nuclear flux(2)$^{\mathrm{a,e,f}}$     \\
  \noalign{\smallskip} 
 \hline 
 \noalign{\smallskip} 
 M~81    &  $u$     &$3465$  &$5.40\pm0.20$   &$4.16\pm0.03$   & $1.2\pm0.6$ & $-6.3\pm0.6$ \\
 	&  $uvw1$  &$2600$  &$4.50\pm0.34$   &$1.06\pm0.03$      &	$3.4\pm0.4$ & $1.5\pm0.4$ \\
 	&  $uvm2$  &$2246$  &$2.33\pm0.25$   &$0.53\pm0.03$     &	$1.8\pm0.4$ & $0.9\pm0.4$ \\
 	&  $uvw2$  &$1928$  &$3.39\pm0.29$   &$0.26\pm0.03$      &	$3.1\pm0.5$ & $2.7\pm0.5$ \\
NGC~3998&  $u$     &$3465$  &$5.73\pm0.02$   &$3.9\pm0.1$ & $1.8\pm0.7$ & $1.0\pm0.7$ \\
        &  $uvw1$  &$2600$  &$3.13\pm0.01$   &$2.4\pm0.1$       & $0.8\pm0.3$  & $0.3\pm0.3$\\
        &  $uvm2$  &$2246$  &$2.46\pm0.01$   &$1.82\pm0.03$  & $0.6\pm0.4$  & $0.3\pm0.4$\\
        &  $uvw2$  &$1928$  &$3.11\pm0.01$   &$1.39\pm0.03$   & $1.7\pm0.4$  & $1.4\pm0.4$\\
NGC~4203&  $u$     &$3465$  &$2.44\pm0.07$   &$2.35\pm0.03$	 & $0.1\pm0.3$  & $-0.4\pm0.3$\\
        &  $uvw1$  &$2600$  &$1.42\pm0.04$   &$0.56\pm0.03$	      & $0.9\pm0.1$ & $0.8\pm0.1$\\
        &  $uvm2$  &$2246$  &$0.62\pm0.02$   &$0.27\pm0.03$  	& $0.3\pm0.1$ & $0.3\pm0.1$\\
        &  $uvw2$  &$1928$  &$0.96\pm0.04$   &$0.13\pm0.03$	     & $0.8\pm0.1$ & $0.8\pm0.1$\\
NGC~4579&  $u$     &$3465$  &$3.11\pm0.06$   &$2.74\pm0.03$	&$0.4\pm0.3$ & $-0.2\pm0.3$\\
        &  $uvw1$  &$2600$  &$1.84\pm0.03$   &$1.33\pm0.03$	& $0.5\pm0.1$ & $0.3\pm0.1$\\
        &  $uvm2$  &$2246$  &$1.55\pm0.06$   &$0.92\pm0.03$	& $0.6\pm0.2$ & $0.4\pm0.2$\\
        &  $uvw2$  &$1928$  &$1.51\pm0.03$   &$0.62\pm0.03$	& $0.9\pm0.2$ & $0.8\pm0.2$\\
 \noalign{\smallskip}
  \hline  
\end{tabular}  
\begin{list}{}{} 
  \item[$^{\mathrm{a}}$]  Not corrected for Galactic reddening, in $10^{-15}$ erg~s$^{-1}$cm$^{-2}$\AA$^{-1}$.
  \item[$^{\mathrm{b}}$]  Total flux, evaluated in a $2\,\arcsec$-radius circular 
area centred on the nucleus, and scaled to the equivalent flux in a $5\,\arcsec$-radius area using the UVOT PSF.  The errors are statistical only (1-$\sigma$).
  \item[$^{\mathrm{c}}$]  Extranuclear galaxy flux  within 
a $2\,\arcsec$-radius circular 
area centred on the nucleus, obtained by interpolating/extrapolating from 
archival HST ACS HRC images at 2500 and 3300\,\AA.  The errors are statistical only  (1-$\sigma$).
  \item[$^{\mathrm{d}}$]  Net nuclear flux, obtained by assuming that the host galaxy radial profile is highly centrally concentrated.
  \item[$^{\mathrm{e}}$]  Net nuclear flux, obtained by assuming that the host galaxy radial profile is nearly flat, i.e. constant within the UVOT $2\,\arcsec$-radius aperture.
   \item[$^{\mathrm{f}}$]  The errors are propagated from the \sw\ UVOT and HST ACS HRC uncertainties and include the systematic uncertainties.

  \end{list} 
  \end{center}
  \end{table*} 


\end{document}